\documentclass{aa}

\usepackage{natbib}
\bibpunct{(}{)}{;}{a}{}{,}
\usepackage{txfonts}

\usepackage{graphicx}

\title{The massive eclipsing LMC Wolf-Rayet binary \object{BAT99-129}. 1\thanks{Based on observations obtained at the New Technology Telescope, La Silla, Chile.}}

\subtitle{Orbital parameters, hydrogen content and spectroscopic characteristics}

\author{C. Foellmi\inst{1} \and  A.F.J. Moffat\inst{2} \and S.V. Marchenko\inst{3}}
\institute{European Southern Observatory, 3107 Alonso de Cordova, casilla 19001, Vitacura, Santiago, Chile
\and D\'epartement de physique, Universit\'e de Montr\'eal, C.P. 6128, Succ. Centre-Ville, Montr\'eal, QC, H3C 3J7, Canada
\and Department of Physics and Astronomy, Thompson Complex Central Wing, Western Kentucky University, Bowling Green, KY 42101-3576, USA}

\offprints{C.Foellmi \email{cfoellmi@eso.org}}

\abstract{\object{BAT99-129} in the LMC is one among a handful of extra-galactic eclipsing Wolf-Rayet binaries known. We present blue, medium-resolution, phase-dependent NTT-EMMI spectra of this system that allow us to separate the spectra of the two components of the binary and to obtain a reliable orbital solution for both stars. We assign an O5V spectral type to the companion, and WN3(h)a to the Wolf-Rayet component. We discuss the spectroscopic characteristics of the system: luminosity ratio, radii, rotation velocities. We find a possible oversynchronous rotation velocity for the O star. Surprisingly, the extracted Wolf-Rayet spectrum clearly shows the presence of blueshifted absorption lines, similar to what has been found in all single hot WN stars in the SMC and some in the LMC. We also discuss the presence of such intrinsic lines in the context of hydrogen in SMC and LMC Wolf-Rayet stars, WR+O binary evolution and GRB progenitors. Altogether, BAT99~129 is the extragalactic counterpart of the well-known Galactic WR binary \object{V444 Cygni}.

}

\date{Received $<$date$>$, Accepted $<$date$>$}

\titlerunning{The massive eclipsing LMC Wolf-Rayet binary BAT99-129. I}

\begin{document}

\maketitle

\section{Introduction}

The Wolf-Rayet (WR) star \object{BAT99-129} \citep[a.k.a Brey 97, see][ hereafter BAT129]{bat99} has been discovered to be a short-period eclipsing binary by \citet*{Foellmi-etal-2003b}. According to these authors, it consists of a nitrogen-rich WN4 Wolf-Rayet component and a companion of unknown spectral-type. Along with \object{BAT99-19} (Brey 16) in the LMC and \object{HD 5980} in the SMC, BAT129 belongs to the very small group of extra-galactic eclipsing Wolf-Rayet binaries. 

Absorption lines are visible in the spectrum of BAT129, especially in the blue part of the optical range ($\lesssim$4000\AA), while WR emission lines strongly dominate everywhere else. The origin of these absorption lines is unclear: they are either produced by an O-type companion, and/or they originate in a WR wind, as has been found in all the {\it single early-type} WN (WNE) stars in the SMC and, to a lesser extent, in the LMC \citep*{Foellmi-etal-2003a,Foellmi-etal-2003b,Foellmi-2004}.

In the SMC about half of the WR population consists of hot, single, hydrogen-containing {\it single} WN3 and WN4 stars (classified WN3-4ha or (h)a). For such hot spectral types, significant amounts of hydrogen were not expected, since such hot WN stars are believed to have peeled off their outer H-rich layers by a very strong and optically thick wind, combined with internal convective mixing to expose core-processed material. Contrary to expectations, blueshifted absorption lines of HI and HeII were detected in the spectra of all single SMC WNE stars, and Foellmi et al. argued that they originate in a WR wind. Such stars were also found in the LMC, and more recently, in our Galaxy, in a region where the ambient metallicity may resemble that of the LMC: WR3, classified as WN3ha \citep{Marchenko-etal-2004}. Finally, two other Galactic, presumably  single WNE stars clearly show hydrogen in their spectra: \object{WR128} (WN4) and \object{WR152} (WN3) \citep{Crowther-etal-1995,Nugis-Niedzielski-1995}.

This paper examines a binary star. Hydrogen is already known to occur in the WR component of some binaries, e.g. in the 1.64-day eclipsing system \object{CQ Cep} \citep{Marchenko-etal-1995}. However, the authors classify the WR component as WN6, while \citet{Smith-1968} gives a spectral type WN7. These two spectral types seem to mostly describe very massive stars \citep[see for instance the most massive star known: \object{WR20a} in our Galaxy, with two stars both of type WN6ha --][]{Bonanos-etal-2004,Rauw-etal-2005} in a core hydrogen-burning transition stage between the O Main Sequence and the core helium-burning WR stage \citep[see also][ for a detailed discussion of this important point]{Foellmi-etal-2003b}. The question that remains to be addressed is: do we detect the presence of hydrogen in the {\it early-type} WN component of a short-period binary, and if yes, could such a presence be explained in the same way?

Previously, none of our spectra of BAT129 were of sufficient quality to detect absorption lines belonging to the companion and measure its radial-velocities (RVs), thus obtaining a reliable orbital solution for both components and helping to separate the two spectra, as well as to resolve the question of hydrogen intrinsic to the WN component.

We present in Section 2 new medium-resolution NTT-EMMI spectra of BAT129. We discuss in detail in Section 3 the iterative procedure we used to separate the spectra and find a reliable orbital solution. In Section 4, we assess the presence of hydrogen in the WR spectrum and examine all the results on this binary star: spectral types of both components, luminosity ratio, radii, mass-loss rates, masses. We discuss in Section 5 the relevance of the presence of hydrogen to the WR classification, WR+O binary evolution and GRB progenitors. We draw conclusions in Section 6.

\section{Observations and Data}

Our spectroscopic dataset of BAT129 is composed of three parts. The first part (binary discovery) was published by \citet{Foellmi-etal-2003b}, to which the reader is referred for a detailed description. There one can also find the MACHO data used to obtain the light curve of BAT129. The second part consists of 21 medium resolution spectra spread over a few days, obtained with the ESO Multi-Mode Instrument (EMMI) at the New Technology Telescope (NTT) on La Silla, Chile. The third part is composed of 12 spectra with identical characteristics to the second dataset, but taken six months later. Table \ref{datasets} summarizes the characteristics of each dataset.

\begin{table*}[!ht]
\caption{(1) \citet{Foellmi-etal-2003b}.  Summary of the 3 spectroscopic datasets obtained on BAT129.}
\label{datasets}
\centering
\begin{tabular}{cccccc} \hline
Dates & Observatory & Instrument & No. of spec. &
Wavel. coverage & Res. power (2.6 pix resol.) \\ \hline
1998-2002 	& See (1) 		& See (1) 	& 24 & 3800-6800\AA	& $\sim$900 	\\
2004, Sep. 22-25, Oct 1st 	& NTT & EMMI	& 21 & 3920-4380\AA	&3500	\\
2005, Mar. 24-26 			& NTT & EMMI	& 12 & 3920-4380\AA	&3500	\\ \hline
\end{tabular}
\end{table*}

\begin{figure}
\centering
\includegraphics[width=\linewidth]{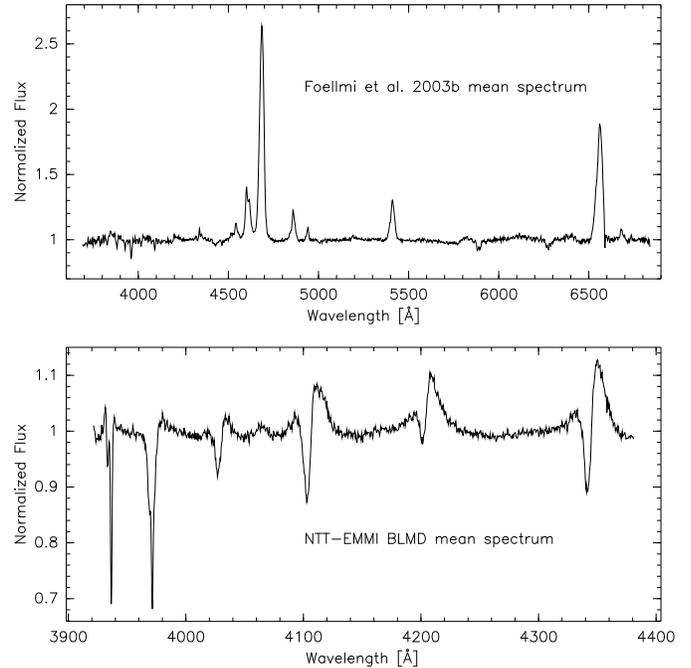}
\caption{Top panel: Mean spectrum of data from Foellmi et al. 2003b. Bottom panel: EMMI mean spectrum. In this spectrum, the emission-line and absorption-line profiles are clearly resolved.}
\label{phd_emmi_spec}
\end{figure}

For the new EMMI data, the Blue Medium-Resolution Spectroscopy mode (BLMD) was used, with Grating\#3 at a central wavelength of 4150 \AA, providing a wavelength coverage of 3920-4380 \AA\ and a dispersion of 0.45 \AA/pixel. The 1"-wide slit provided a resolving power of 3500 (resolution of 2.6 pixels). The typical 30-min exposure resulted in S/N ratio between 50 and 90. The spectra were reduced using standard IRAF tasks and shifted to the heliocentric rest frame. Using continuum regions common to all EMMI spectra, the continuum was normalized to unity. The mean spectra of dataset~1, and datasets~2 and~3 together, are shown in Fig.~\ref{phd_emmi_spec}. It can be seen that the new spectra allow us to clearly resolve the emission and absorption components.

\section{Analysis}

We performed an iterative analysis on the EMMI spectra, that leads to a new and reliable orbital solution as well as separation of the individual spectra of the components. The general appearance of the  extracted spectra, and therefore the shapes and strengths of features found in them, are very sensitive to the process of separation. We describe below in detail the analysis and the decisions we took in order to reach a reliable result.

\subsection{Photometric period, ephemeris and continuum correction}

Before computing the orbital phases for the EMMI spectra, we recalculated the ephemeris (period and zero point $E_0$) published by \citet{Foellmi-etal-2003b}. The published values were based on the orbital solution obtained for RVs of emission lines only. In our new approach, we tried to minimize any influence of wind-wind collision (WWC) effects on the RVs. Furthermore, emission-line RVs of WR stars may not be a good indicator of  the systemic RV, although orbital motion may not be a problem per se. This point is also discussed below.

\begin{figure}
\centering
\includegraphics[width=\linewidth]{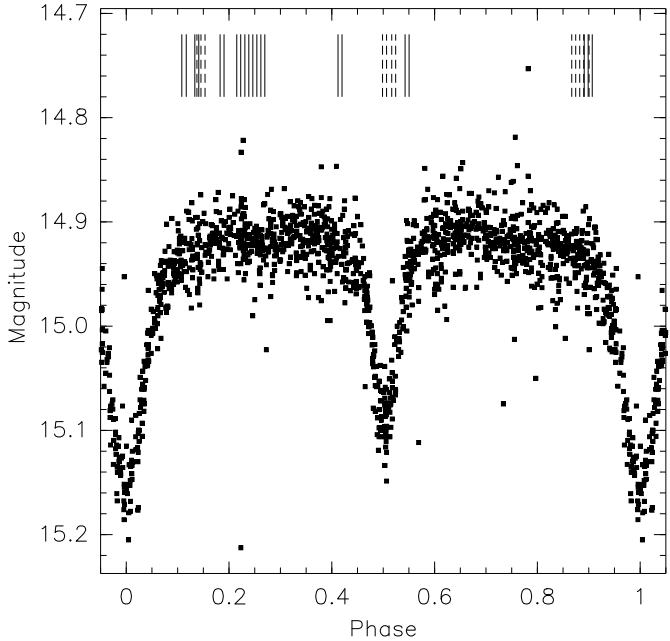}
\caption{MACHO light curve of BAT129, as described in \citet{Foellmi-etal-2003b}, with the new value of $E_0$. MACHO instrumental magnitudes were transformed to obtain a mean magnitude outside the eclipse of 14.92 \citep{bat99}. The phases at which EMMI spectra were obtained are indicated by vertical lines. Full lines indicate spectra from the second dataset, whereas dashed lines indicate spectra from the third dataset.}
\label{phot_corr}
\end{figure}

A period search in the combined R and B MACHO data led us to a period of 2.7689 $\pm$ 0.0002 days, which is equivalent to the published value. We note that with the I-band OGLE data \citep[ID: LMC\_SC18 122046/OGLE054148.68-703531.0; see also][]{Foellmi-etal-2003b}, \citet{Wyrzykowski-etal-2003} obtain a value of 2.76895 days. To determine the zero point (i.e. time of primary minimum) $E_0$, we applied the bisector method on the deeper eclipse light-curve minimum, and obtained the following value: $E_0$ [HJD] = 2448843.8935 $\pm$ 0.0007. The resulting light curve folded in phase is shown in
Fig.~\ref{phot_corr}.

In order to be used in the separation procedure, the EMMI spectra must be corrected for the variable continuum level. We smoothed the lightcurve by taking mean values from 0.01 phase bins, converted magnitudes into fluxes and interpolated the smoothed lightcurve at the appropriate phases of the RV data. Then we computed the ratio between the flux at a given phase and the mean flux taken outside the eclipses, and appropriately normalized each spectrum by this ratio. The correction was at most 15\%. In that manner, the shapes (and equivalent widths) of the spectral lines do not depend on the variable continuum level.

\subsection{The "old" radial-velocities}
\label{old-rvs}

We also decided to re-measure the RVs of the first dataset, clearly affected by the traveling WWC emission details. New measurements were made using the bisector method on the strongest and best defined line: HeII~$\lambda$4686.\footnote{We also assessed the HeII~$\lambda$5411 line, but discarded the measurements because of the relatively low S/N ratio in $\sim 50$\% of the spectra.} The bottom level for the bisectors was fixed at 1.125 (the continuum being at unity), in order to avoid blending problems. To minimize the influence of the emission component arising in the WWC zone, we fixed the top level at 1.5, thus using the lower-third portion of the line. This proved to be quite reliable, since the results did not change significantly if we slightly modified these limits.

\subsection{The preliminary radial-velocities of the absorption profiles}

Before measuring  RVs in the EMMI spectra, we used the two narrow interstellar Calcium lines H \& K at 3933 and 3968 \AA, clearly visible in the EMMI spectra, to self-calibrate the RVs by co-aligning the IS features in the individual spectra, in RV space. The correction was at most 30 km/s for a given spectrum. We then removed these lines using gaussian fitting, in order to avoid problems during the separation process (see below).

Simultaneous measurement of the emission and absorption profiles in the original EMMI spectra proved to be problematic. The uncertainties are introduced both by (a) the large-amplitude relative motion of the absorption and emission profiles, and (b) the traveling emission component produced in the WWC zone. We therefore chose at that point to measure the absorption profiles only, since they are relatively well-defined in each individual spectrum.

We measured positions of the absorption profiles via least-square fitting of Gaussian profiles (the IRAF task "splot") for the following lines: HeII~$\lambda$4025, H$\delta$, HeII~$\lambda$4200 and H$\gamma$. A weight of 0.5 was assigned to the relatively weaker absorptions at $\lambda$4025 and $\lambda$4200, compared to 1.0 for the rest. We thus combined these four lines into a single set of absorption-line RVs.\footnote{The situation is slightly more complicated since HeII~$\lambda$4025 is in fact a blend of the lines HeII~$\lambda$4025.60 and HeI~$\lambda$4026.191. Moreover, the Balmer lines H$\delta$ and H$\gamma$ are blended with HeII lines as well. In order to be perfectly precise, the positions of the profiles measured on the spectra should be compared to the rest wavelength of the corresponding lines. But in the case of blends, the contribution of each individual line is not known. We make here the reasonable hypothesis that the relative intensity contribution of one individual line to a given blend is not changing with time. Therefore, the comparison of a profile position to one rest wavelength is relevant and coherent. Thus the choice of one rest wavelength to another will only produce an offset in the absolute value of the RVs. At that point of our analysis, only relative RVs are meaningful. In general, except when specifically stated, the rest wavelength that is used is the one of the line specified explicitly in the text.}

\subsection{The first-iteration orbital solution}
\label{first_solution}

Combination of the newly obtained  RVs of the emission lines (dataset 1) with the RVs of the absorption lines in EMMI spectra leads to a complete set of spectroscopic orbital parameters. We applied the algorithm described in detail by \citet{Bertiau-Grobben-1969} to obtain the solution. We fixed the period to the photometric value. Since the derived eccentricity did not significantly differ from zero, we subsequently adopted e = 0.0 (but see below).

On the other hand, we allowed the emission-line RVs to exhibit a phase shift.  Although our bisector measurements aimed to avoid (at least, substantially minimize) any WWC influence, it is still possible that these RVs may show residual WWC or other effects, therefore misrepresenting the genuine orbital motion \citep[see the detailed discussion in ][]{Lewis-etal-1993}. We allowed the phase shift to be an additional free parameter, introducing it in discrete steps of 0.001 days, from -0.2 to 0.2 days. We thus computed a series of orbital solutions, artificially shifting the time of emission-line RVs with these step values, and selecting the solution with the minimal rms deviation between the observed and calculated RVs. Table \ref{first_orbital_solution} shows the comparison between this solution and that previously published \citep{Foellmi-etal-2003b}.

Clearly, the newly obtained RVs provide much smaller K$_{WR}$ amplitude, by about 80~km~s$^{-1}$. This is due to the newer, better emission-line RVs from dataset~1, and resulting higher-quality fit of our RVs. The phase shift between the two sets of RVs is equal to 0.012 days, i.e. 17 minutes, which is negligible. Moreover, the maximum RV difference between this solution and a solution with a null phase shift does not exceed 5 km~s$^{-1}$. The two sets of RVs can therefore be considered as being well in phase with each other.

\begin{table}[t]
\caption{First iteration of orbital parameters of BAT129. The parameters published by Foellmi et al. 2003b are also shown for comparison.}
\label{first_orbital_solution}
\centering
\begin{tabular}{lll} 
\hline \hline
Parameter & Foellmi et al. 2003b & This study \\ \hline
P (days)						& 2.7687 $\pm$ 0.0002	& 2.7689 (fixed)  \\
K$_{WR}$ (km\,s$^{-1}$)		& 398 $\pm$ 7	 		& 318 $\pm$ 7 \\
K$_{O}$ (km\,s$^{-1}$)		& - 						& 174 $\pm$ 7 \\
e  							& 0.082 $\pm$ 0.022  	& 0. (fixed) \\
E$_{0}$ (HJD- 2 451 100.0) 	& 820.47 $\pm$ 0.09	& 846.4468 $\pm$ 0.009 \\
$\gamma$	 (km\,s$^{-1}$)		& 245 $\pm$ 15 			& 270 $\pm$ 5 \\
$\omega$	 ($^\circ$)			& 342 $\pm$ 12 			& 270 (fixed)\\
a$_{WR}$ sin~i ($R_{\odot}$)		& - 						& 17.5 $\pm$ 0.5 \\
a$_{O}$ sin~i ($R_{\odot}$)		& - 						& 9.5  $\pm$ 0.6 \\ 
$\sigma$(O-C)$_{WR}$ (km\,s$^{-1}$) &		-			& 39 \\ 
$\sigma$(O-C)$_{O}$ (km\,s$^{-1}$) &		-	& 17 \\ 
\hline
\end{tabular}
\end{table}

\subsection{The first attempt at spectral separation}

We used the calculated RVs in an attempt to separate the spectral components in
the EMMI data following the iterative method from \citet{Demers-etal-2002} \citep[See also][]{Marchenko-etal-1998}. This method consists of the following steps, starting with the emission-line component:

\begin{enumerate}
\item Shift the spectra (always in RV space) to the common rest frame of the WR star, using the (fitted) orbital RVs of emission lines,
\item Combine the shifted spectra (possibly weighted by the individual S/N ratios), thus providing a "super WR spectrum", where the O-star features have been smeared out.
\item Shift the resulting mean spectrum back to the original positions, one by one.
\item Subtract the mean spectrum from the {\it original} spectra, thus obtaining a set of first-iteration, individual spectra of the O component. 
\item Using the RVs of absorption lines, shift these first-iteration absorption-line spectra to the common rest frame of the O star.
\item Combine them (possibly weighted by individual S/N ratios), thus providing a mean first-iteration absorption-line spectrum.
\item Shift the mean absorption-line spectrum back to the original positions, one by one (either using the previously calculated RVs or estimating them via cross-correlation).
\item Subtract this mean absorption spectrum from each of the {\it original} spectra, thus providing a better guess at the emission-line spectrum and finishing the iteration.
\end{enumerate}

The process is repeated until a stable solution is found, usually after 3 to 5 iterations. The choices to be made, as well as sources of uncertainty in the separation procedure are the following:
\begin{itemize}
\item The weights during the combination of spectra.
\item The possibility of editing by hand the combined spectra.
\item The possibility to compute (via Cross-Correlation = CC) new RVs in the currently extracted spectra, in each iteration, before shifting.
\item The wavelength range used in the CC.
\item The reference spectrum used in the CC.
\item The spectra to be combined into a mean spectrum (all phases, undiscretionately or some only at specific phases?).
\end{itemize}

After many tests, we converged to the following: 1. We decided not to use weights when combining the spectra in order not to favor any orbital phase, as we are dealing with a relatively small number of spectra. Most of our EMMI spectra have about the same S/N ratio (between 50 and 90). This option proved to be of negligible influence. 2. Editing by hand the combined spectra proved to be important in order to force the process to converge rapidly. The editing was done in order to minimize the adverse alteration of the spectrum, by simply erasing absorption profiles from the WR spectrum and emission details (presumably related to the WWC effects) from the absorption spectrum. Tests reveal that this is absolutely necessary for adequate convergence. 3. When using the CC for computing new RVs before shifting, we chose the wavelength range 4000-4370\AA, i.e. the spectral range comprising the 4 most important absorption lines: HeII~$\lambda$4025, H$\delta$, HeII~$\lambda$4200 and H$\gamma$. 4. The reference spectrum used in the CC computations was chosen to be the average spectrum from the previous iteration. 5. Finally, the spectra around phase 0.9 show very clear WWC emission line excess, and their profiles considerably influence the resulting mean spectrum. They are certainly not representative of an unperturbed WR spectrum, and their influence must be tested.

In order to test our separation process, we performed 4 different separations, called hereafter "loops". In the first two loops we used the fitted RVs from the orbital solution above during the first 2 iterations, then switching to the RVs computed with CC in 3 additional iterations. In the first loop we did not take the spectra at phase 0.9 when compiling the averages. Note that the 0.9-phase spectra were kept in the process, and we obtained extracted components at those phases, but they were not used during the combination of the spectra (steps 2 and 6). In the second loop we used all the spectra.

The two other loops used the RVs from the orbital solution only, during 5 iterations. No new RVs were computed with the CC method, thus final results being a 'black-box' output of the iterative process, based on preliminary RVs of the first orbital solution. Again, we removed the phase 0.9 spectra when calculating the mean spectrum in the third loop, and treated  all phases equally in the fourth loop.

The resulting spectra of the 4 loops were found to be very similar, at first sight. However, significant differences were found in the RVs that were computed from them. In fact, since we don't know {\it a priori} the correct shape of the spectrum of the WR and the O components, we have to find a way to discriminate between these 4 approaches. We used orbital solutions to achieve this.

\subsection{Exploring the parameter space of the orbital solutions}

As stated in the first orbital solution above, the emission-line RVs from dataset 1, and the absorption-line RVs from EMMI spectra seem to be well in phase. To check this point again with the new absorption-line RVs on extracted O-star spectra, for each of the 4 different loops, we computed the RVs for the O spectra via CC in the wavelength range 4000-4370~\AA. Then, we looked for an orbital solution taking only these RVs and the emission-line RVs of dataset-1, allowing for a phase shift (i.e., again taking the solution with the smallest $\sigma (O-C)$) as described above. We found 4 solutions in quite good agreement with the first orbital solution (see section \ref{first_solution}), with phase shifts ranging from -0.067 days (-1.6 hour, 2\% of the period) to -0.012 days (0.3 hours, 0.4\% of the period). This can be regarded as negligible, meaning that these EMMI {\it absorption-line} RVs obtained on extracted spectra are well in phase with the RVs of dataset-1, and give a stable overall result. These two sets of RVs can therefore be combined into a single set, and give us confidence on the results of the separation algorithm.

Second, we looked for orbital solutions taking the three sets of RVs, allowing for a phase shift of the emission-line RVs from the extracted WR spectra {\it only} (since the previous 2 datasets were found well in phase). Moreover, we assume here that the number of spectra we have and the coverage in phase they provide are good enough to find a reliable solution, without additional constraints such as a circular orbit. We assume that the best solution is the one which gives the minimal scatter between fitted and observed RVs, {\it and} the smallest value of the eccentricity {\it and} the smallest phase shift. For each of the 4 different loops, we looked for an orbital solution with a free phase shift between -0.2 and 0.2 day in steps of 0.001 day and with eccentricity as a free parameter.

For each of the 4 loops, we have two "best solutions", one with the smallest scatter $\sigma (O-C)$ and one with the smallest eccentricity. It appears clearly in our results that the third loop (i.e. blind separation of 5 iterations without combining the 0.9-phase spectra) provided both smallest values for $\sigma (O-C)$ (of 36.6 and 39.2 km\,s$^{-1}$, respectively), while all other solutions have a $\sigma (O-C)$ above 40 km\,s$^{-1}$. The set of spectra providing the orbital solutions of the best quality is therefore that of the third loop.

\begin{figure}[!ht]
\centering
\includegraphics[width=\linewidth]{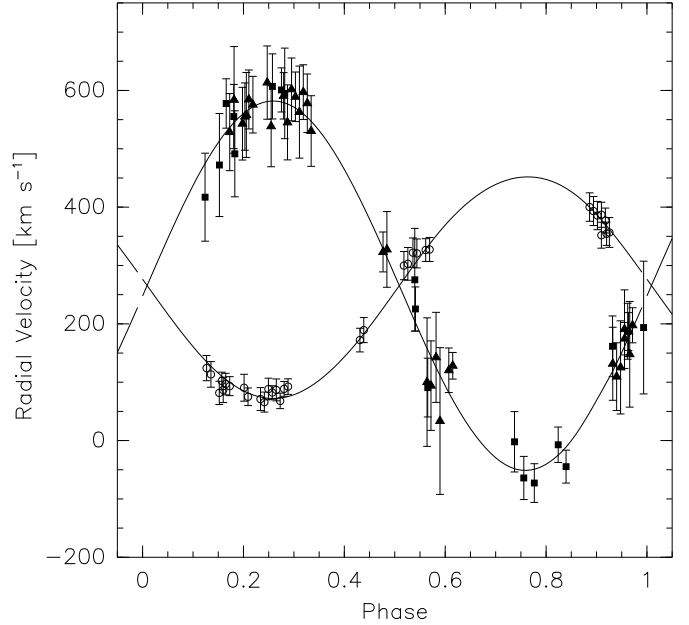}
\caption{Final (adopted) circular orbital solution including all 3 sets of RVs. Squares: RVs of emission lines (dataset 1). Triangles: RVs of emission lines (datasets 2 and 3). Circles: RVs of absorption lines (datasets 2 and 3).}
\label{finalorbit-fig}
\end{figure}

\begin{table}[!ht]
\caption{Adopted circular orbit of BAT129.}
\label{finalorbit-tab}
\centering
\begin{tabular}{ll} 
\hline \hline
Parameter			& Circular orbit (adopted) \\ \hline
P (days)			& 2.7689 (fixed) \\
K$_{WR}$ (km\,s$^{-1}$)		& 316 $\pm$ 5 \\
K$_{O}$ (km\,s$^{-1}$)		& 193 $\pm$ 6\\
e  							& 0. (fixed) \\
E$_{0}$ 					& 2451946.4274 $\pm$ 0.0054 \\
$\gamma$ (km\,s$^{-1}$)		& 265 $\pm$ 5 \\
a$_{WR}$ sin~i ($R_{\odot}$)			& 17.3 $\pm$ 0.4 \\
a$_{O}$ sin~i ($R_{\odot}$)				& 10.6 $\pm$ 0.5 \\
$\sigma$(O-C)$_{WR}$ (km\,s$^{-1}$) 	& 34.0 \\
$\sigma$(O-C)$_{O}$ (km\,s$^{-1}$) 		& 13.4 \\
M$_{WR}$ sin$^3$~i ($M_{\odot}$)		& 14 $\pm$ 2 \\
M$_{O}$ sin$^3$~i ($M_{\odot}$)	 		& 23 $\pm$ 2 \\ 
\hline
\end{tabular}
\end{table}

It is not too surprising that a "blind" separation loop produces the best results. Although 2 blind iterations (in loops 1 and 2) allow the process to converge rapidly to mean WR and O profiles, the next iterations with RVs computed from them are still very sensitive to any residuals.

When comparing the two "best" solutions of the third loop, it is clear that they are very similar, and can be considered as equivalent. All orbital parameters are identical within the errors, except for the phase shift. The solution with the smallest $\sigma$ has a phase shift of 0.127 days (3.05 hours), while the one with the smallest eccentricity has a phase shift of 0.076 days (1.8 hours). Given the fact that both solutions have a resulting eccentricity consistent with an almost circular orbit: 0.022 and 0.021 $\pm$ 0.015, respectively, we decided to choose the solution with the smallest $\sigma$ (i.e. the one with the 0.127-day phase shift), and enforce a circular orbit for simplicity. Finally we adopted the circular orbit, summarized in Table \ref{finalorbit-tab}, and shown in Fig.~\ref{finalorbit-fig}.

\subsection{The final separation}
\label{final-sep}

\begin{figure}[!ht]
\centering
\includegraphics[width=\linewidth]{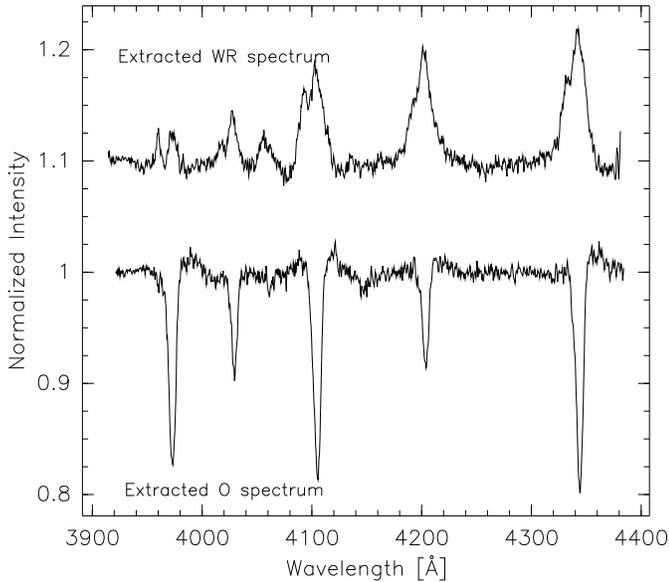}
\caption{Individual spectra of the WR and O components. The WR spectrum has been shifted vertically by 0.1, while keeping the same scale.}
\label{blmd_separation_final}
\end{figure}

Using the fitted RVs from the last orbital solution, we performed an ultimate 'blind' (no CC measurements) separation with 5 iterations, with a limited set of spectra to combine (i.e., excluding phase 0.9) and no weights applied during the averaging. The extracted spectra are shown in Fig.~\ref{blmd_separation_final}.

It must be noted here that the shapes of the wings of absorption and emission lines are very sensitive to the adopted levels of local continuum. Given the broadness of the emission lines and their substantial shifts due to orbital motion, it is difficult to define adequately wide {\it common} line-free regions to be used during continuum rectification. The most apparent consequence of a non-optimal continuum fit is the presence of low-intensity emission excesses in the wings of absorption lines in the O-star spectrum (see Fig.~\ref{blmd_separation_final}). To alleviate this problem, we repeated the whole separation procedure (i.e. from section \ref{old-rvs} to \ref{final-sep}) {\it twice}, changing the approach to the continuum rectification. The continuum fits were done with the same polynomial, but once with a maximal number of available continuum regions and then with a minimal number of samples. The further steps were performed identically and led to the same sequence of decisions between the different orbital solutions. Finally, we chose the continuum rectification  that provides the best (i.e. flattest) continuum on the  final, extracted spectra, i.e. the one with the maximum continuum samples.

\section{Results}

We discuss here the results of the separation loops, i.e. the resulting separated spectra, their properties, and the orbital solution.

\subsection{The Wolf-Rayet absorption features}

One important detail can be seen in the separated WR spectra: absorption profiles on the blue sides of the emission lines. These features were always present in the final results of our separation tests. Hence, the presence of absorptions does not depend on our analysis of the continuum or on the editing 'style' of the mean spectrum during the separation, nor does it depend on the choice of  iterative sequence  and the system of RVs we adopt while performing the separation.

\begin{figure}
\centering
\includegraphics[width=0.8\linewidth]{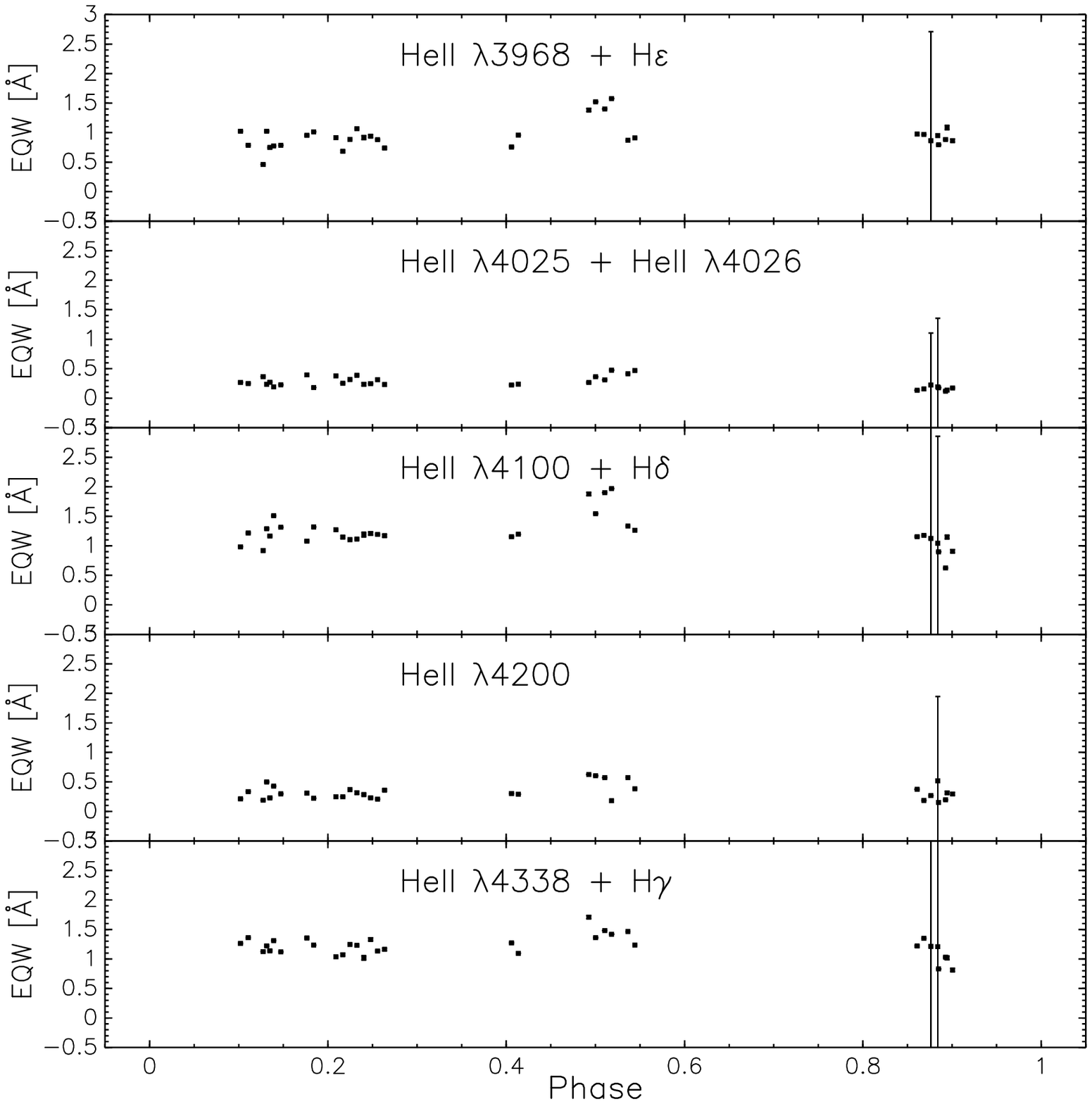}\\
\includegraphics[width=0.8\linewidth]{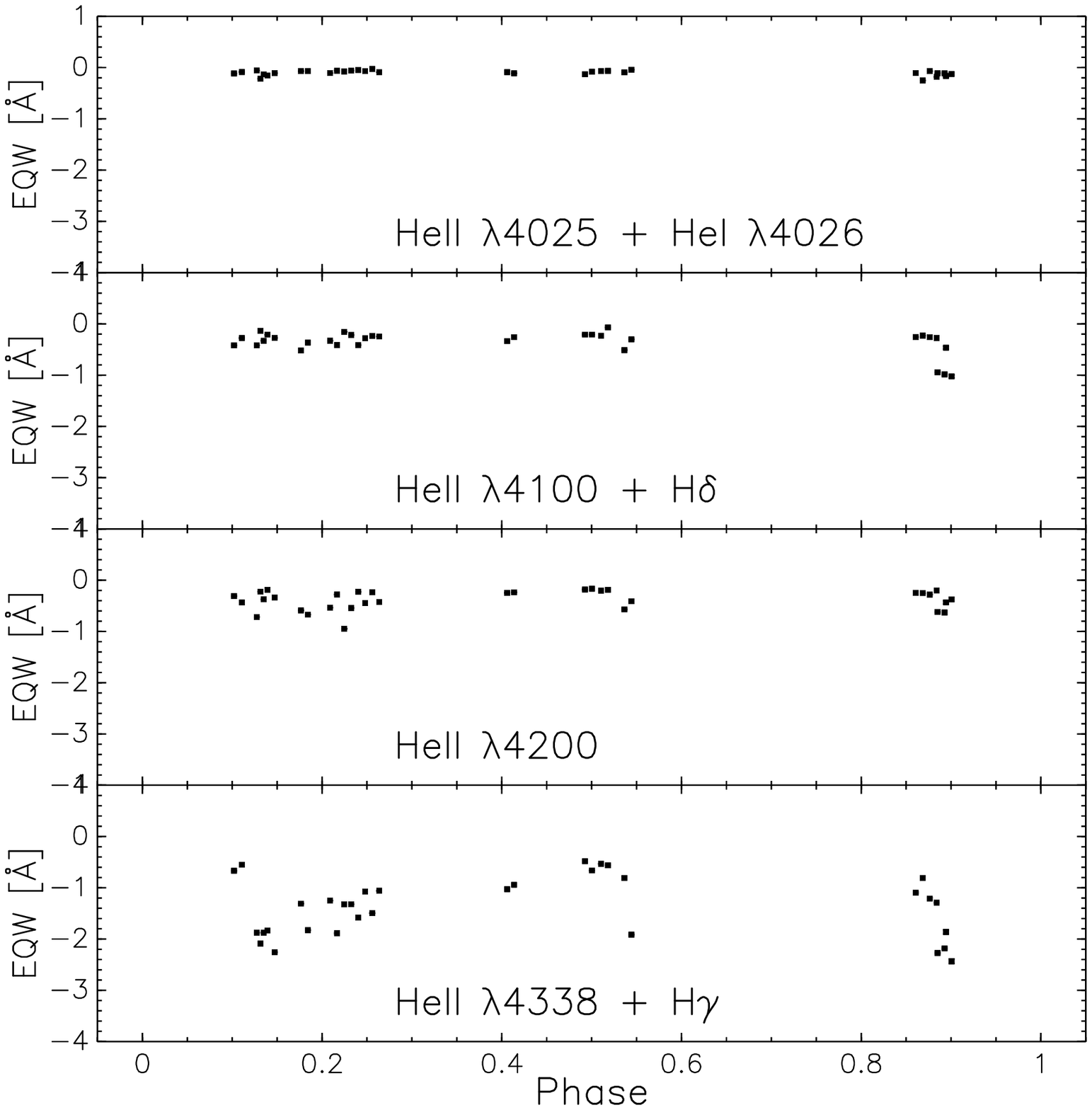}
\caption{Top figure: EW of absorption lines measured on the extracted O-star spectra. Bottom figure: EW of emission lines measured on the extracted WR spectra, including the WR absorption features. Note that panels of the two figures do not have the same vertical scale.}
\label{eqws}
\end{figure}

Despite what is stated above, and in order to strengthen our confidence in the result of the separation, we measured the equivalent widths (EWs) of the 5 absorption lines (i.e. including this time H$\epsilon$) seen in the O spectrum. If the separation process is reliable, we might expect that these EWs should not depend on the orbital phase. We also measured the EW of the 4 most prominent emission lines in the WR spectrum. The results are  shown in Fig.~\ref{eqws}. It appears that the EWs of the absorption lines in the individual extracted O-star spectra are roughly constant. No obvious periodic pattern or phase-locked variations are detectable in these EWs. Moreover, the EWs of the emission lines in the WR spectra are also relatively constant (see Fig.~\ref{eqws} - bottom figure).

Continuing to test the sensitivity of the overall appearance of the weak absorption components to details of the extraction algorithm, we attempted to construct an unperturbed WR spectrum. For that, we shifted the WR spectra to the same reference frame using the RVs of the accepted orbital solution. Then, we produced an average from the spectra around phases 0.4-0.6 (corresponding to inferior conjunction of the O star). The red side (up to 2000 km~s$^{-1}$) of the profiles was assumed to represent an unperturbed profile. Then, we combined the 3 spectra around phase 0.15 and considered the blue side of the average (down to -2000 km~s$^{-1}$) as a fair representation of the unperturbed WR line profile. Then we merged the two sides for each line and subtracted the templates from the individual spectra. Unfortunately, the relatively low S/N ratio of the spectra disallowed any further analysis of the absorption features, which were submerged in noise in all individual spectra.

\begin{figure*}
\centering
\includegraphics[width=0.25\linewidth]{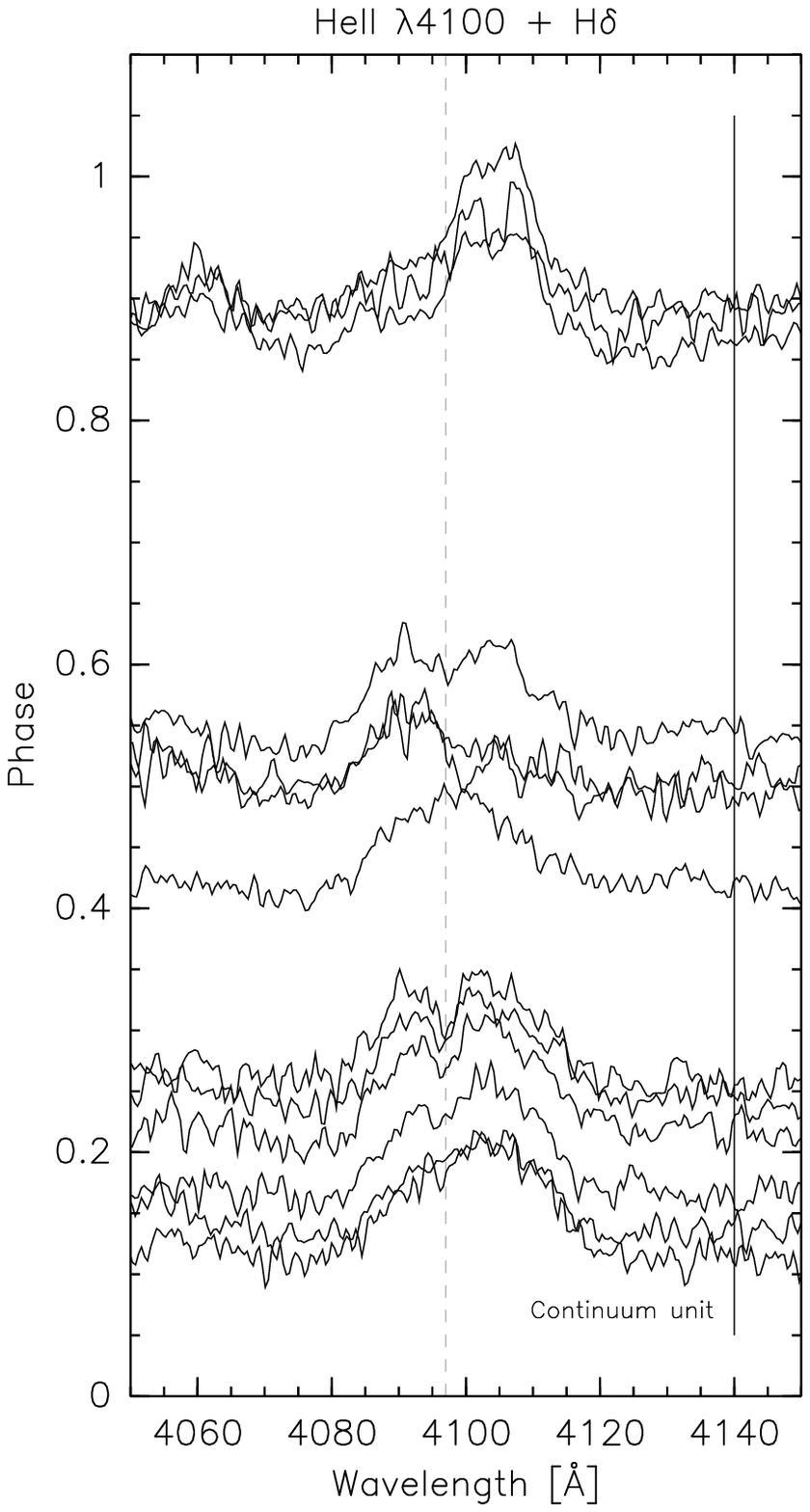}\includegraphics[width=0.25\linewidth]{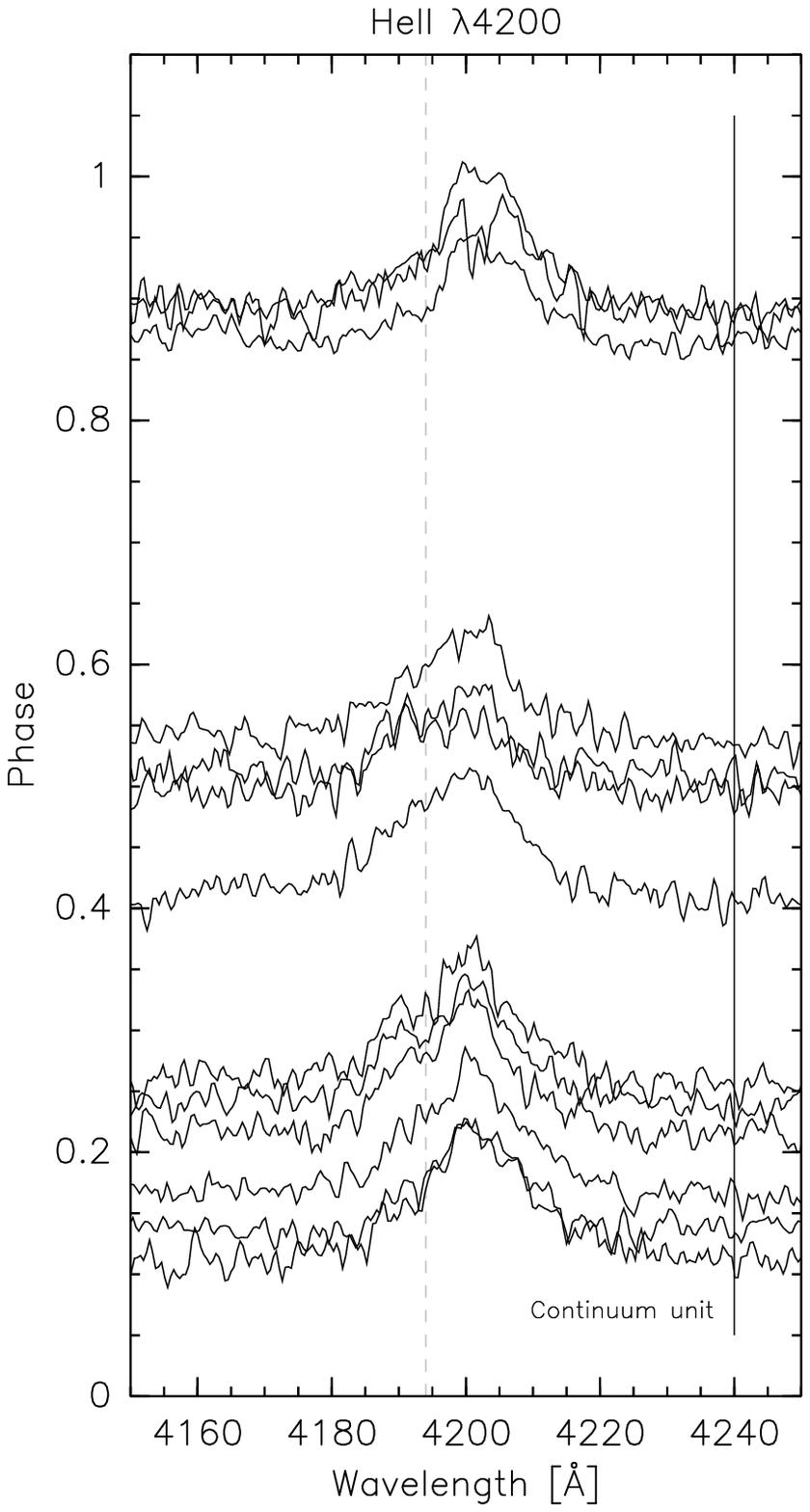}\includegraphics[width=0.25\linewidth]{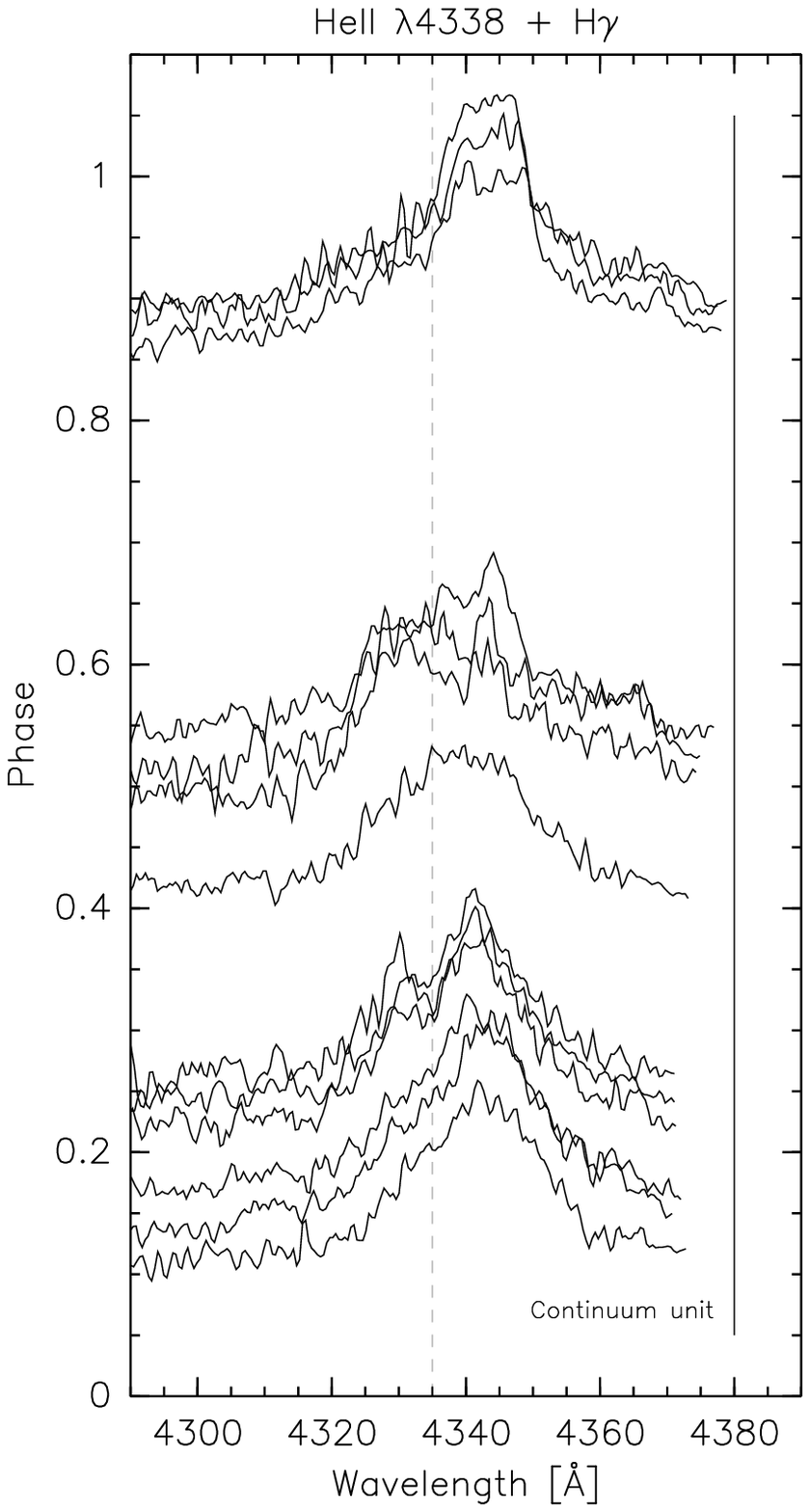}\includegraphics[width=0.25\linewidth]{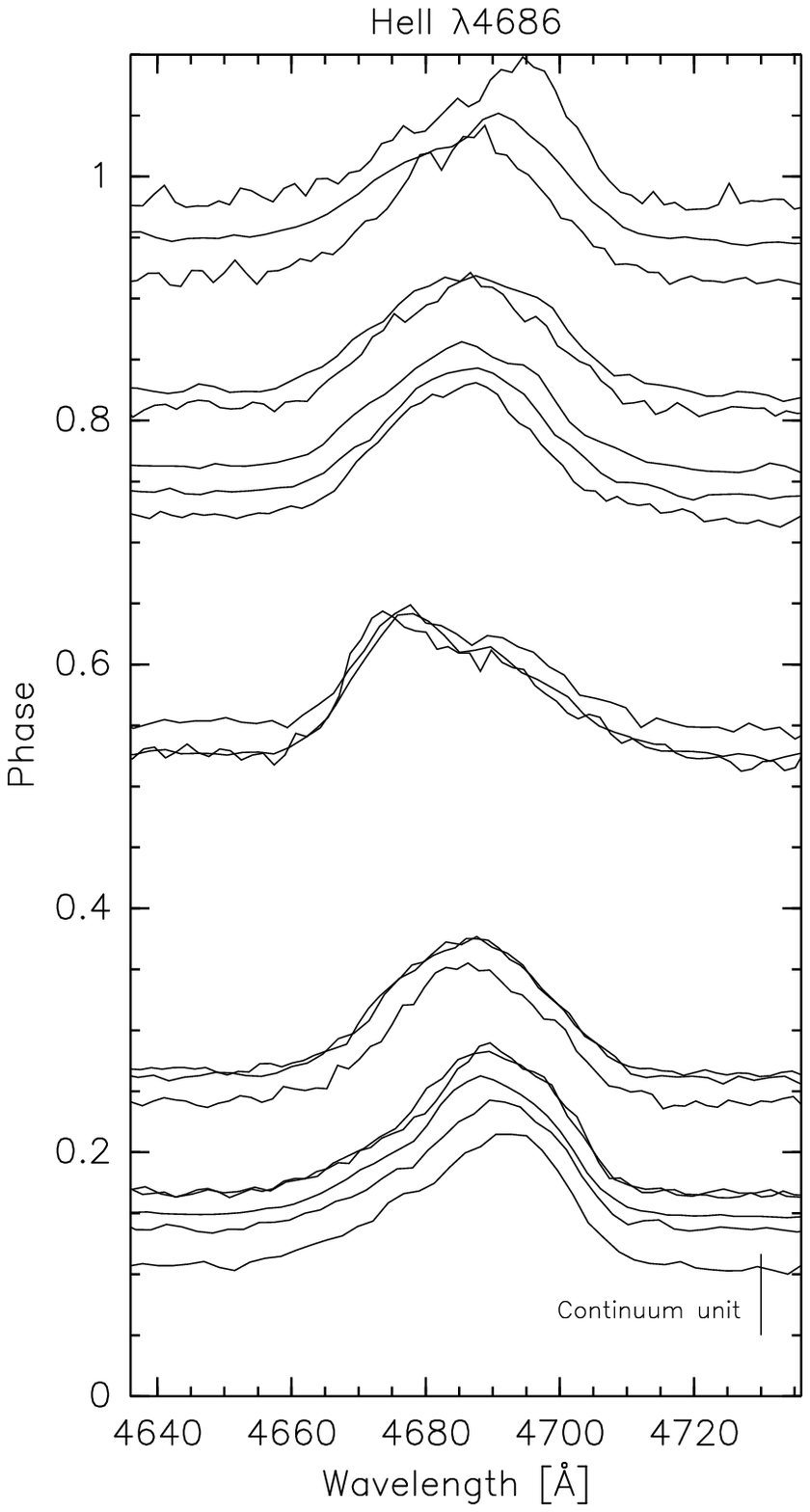}
\caption{From left to right: Montage of H$\delta$, HeII4200, H$\gamma$ and HeII4686 from the WR component plus the WWC emission. All spectra were shifted to the WR rest-frame. For HeII4686 we used a reduction scale factor of 15. In the first 3 figures, the mean position of the absorption line is indicated with a dashed gray line.}
\label{grays}
\end{figure*}

A montage of the most prominent lines in the WR rest frame can also be used to understand more precisely what happens. The plots of H$\delta$, HeII~$\lambda$4200, H$\gamma$ (all coming from the EMMI data) and HeII~$\lambda$4686 (our dataset-1) are shown in Fig.~\ref{grays}. The montage of the HeII~$\lambda$4686 profiles shows very clearly the WWC excess-emission, traveling from the red side (around phase zero) when the O star is behind and the shock cone is pointing away from the observer, and reaching its bluemost position around phase 0.5. On the montages of the three other lines, it can be seen that the absorption feature is most clearly seen between phases 0.15 and 0.25. This is consistent with the fact that an excess emission is present in the red-center portion of the profile during that narrow phase interval. Then, when traveling to the blue, around phase 0.5, this emission excess fills in the absorption feature, then retreating to the red side at phases around 0.9 and producing a profile with a pronounced redshifted bulge. As mentioned before, the latter caused problems during the separation process.

\begin{figure}[!ht]
\centering
\includegraphics[width=\linewidth]{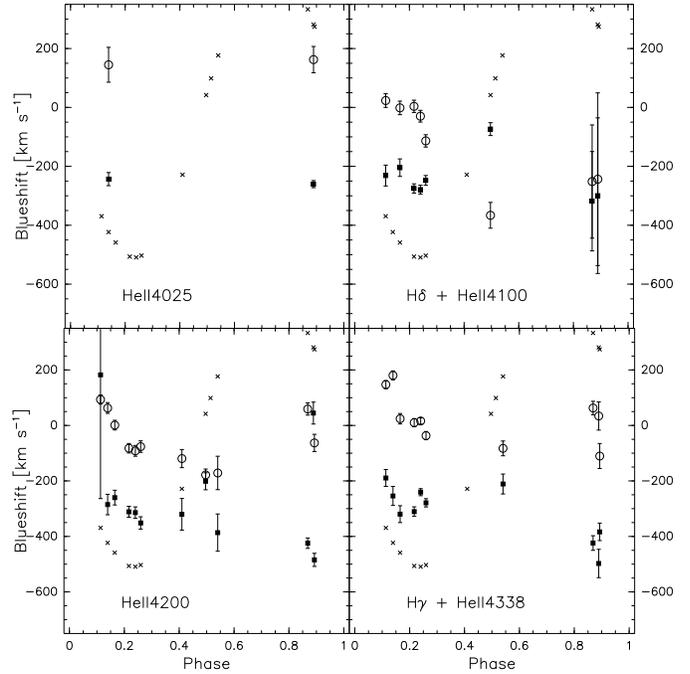}
\caption{Velocities of the WR absorption and emission lines obtained from the positions of the double gaussian fits, versus orbital phase. The positions are compared to the rest wavelength of the corresponding line. For the H+He blends, the simple average of the rest wavelengths has been taken. Filled squares: absorption lines. Open circles: emission lines. The crosses indicate the orbital RVs of the O-star in the WR frame.}
\label{blueshifts}
\end{figure}

The ultimate test to check if the WR absorption lines do belong to the WR star is to measure the blueshifts of these absorptions, assessing both the blueshift on the mean WR spectrum (that is a measure of the separation between the WR absorptions and emission lines) and also blueshifts seen in individual lines (to prove that the
blue-shifted absorption lines follow the WR orbital motion). To measure the blueshifts on individual spectra, we first shifted the WR spectra to the same frame using the orbital RVs, then combined them in orbital phase bins of 2 or 3 spectra in order to increase the S/N ratio. A double gaussian fit was performed on the 4 prominent lines cited above: HeII~$\lambda$4025, H$\delta$, HeII~$\lambda$4200 and H$\gamma$, hence providing a position both for the absorption and the emission profile. The results are displayed in Fig.~\ref{blueshifts} (including only the measurements for which the fit provided a reasonable solution). The squares show the velocities of the absorption lines compared to the rest wavelength. The open circles show the velocities of the emission lines.\footnote{For the H+He blends, the simple average of the rest wavelength has been taken. The important point here is to compare the velocities of the absorption and emission profiles together (see also previous footnote).}

Velocities of absorption and emission lines in Fig.~\ref{blueshifts} show a rough trend to be slightly in anti-phase. This is particularly clear for H$\gamma$ where initially the RVs of emission and absorption are well separated, then become almost identical at phase 0.5, and again move apart at phase 0.9. Recall that all measurements are made in the WR rest-frame. Clearly, the excess emission coming from the WWC dominates the fits, apparently shifting the absorption line. However, the amplitude of the shifts of the absorption lines is much smaller than the corresponding O-star RVs, whose amplitude would be "enlarged" in the WR rest frame (the O-star motion is indicated by crosses in Fig.~\ref{blueshifts}), and thus the absorption lines cannot belong to the O component. Disregarding the WWC-induced drift, the blueshift of the absorption lines can therefore be considered as constant relative to the rest wavelength, in the WR frame.

Finally, we took all the shifted WR spectra except those at phase 0.9 and combined them into a simple average, to increase the S/N ratio. We then performed a double-gaussian fit on the 4 prominent lines. The results are summarized in Table \ref{mean_blues}, where the positions of the absorption and emission lines are compared {\it to each other}. The blueshift is clearly increasing with lower Balmer/Pickering number(s).

\begin{table}[!ht]
\caption{Blueshifts of WR absorption lines when compared to the positions of the emission lines.}
\label{mean_blues}
\centering
\begin{tabular}{lr}
\hline \hline
Line(s) & Blueshift (km~s$^{-1}$) \\ \hline
HeI~$\lambda$4025 + HeII~$\lambda$4026 	&  -86.3 $\pm$ 18.3 \\
H$\delta$ + HeII~$\lambda$4100 	& -161.8 $\pm$ 12.5 \\
HeII~$\lambda$4200				& -228.7 $\pm$ 11.5 \\
H$\gamma$ + HeII~$\lambda$4338	& -289.0 $\pm$ 10.6 \\ 
\hline
\end{tabular}
\end{table}

Altogether, these tests on the absorption lines seen in the WR spectrum allow us to claim that these blue-shifted lines belong to the WR star itself. Therefore, they are probably formed in the wind of the WR star, similarly to what has been seen in other single SMC and LMC stars \citep[see][ in particular]{Foellmi-etal-2003a}.

\subsection{Spectral types of the WR and O stars}

\begin{figure}[!ht]
\centering
\includegraphics[width=\linewidth]{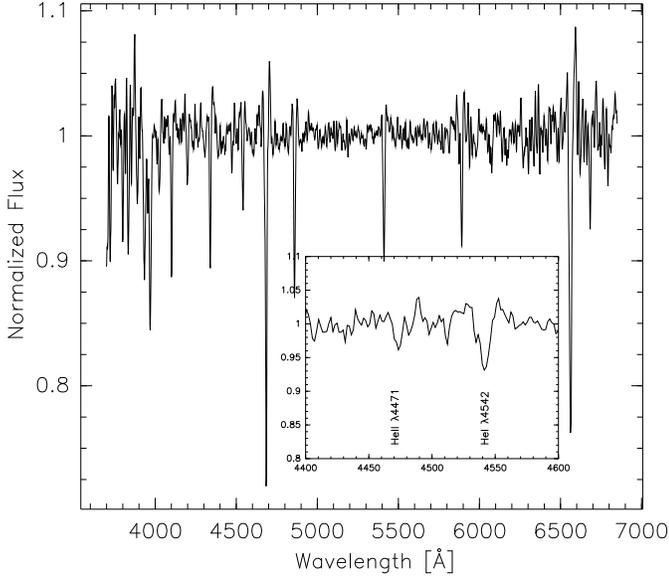}
\caption{O-star spectrum obtained by a separation process on the spectra of dataset 1. The inset shows the region around the two lines HeII~$\lambda$4471 and HeI~$\lambda$4542 used for classification.}
\label{phd_Ostar}
\end{figure}

Although they have a lower quality and the dataset is not homogeneous, we attempted to separate the orbital components with the spectra of dataset 1, using a blind separation with 5 iterations, the orbital RVs and no editing. Thanks to the broader wavelength coverage and the lower resolution, we obtained a reasonably good O-star spectrum. The result is displayed in Fig.~\ref{phd_Ostar}. A comparison of the mean O-star spectrum in \object{BAT99~129} with standard spectra in the same (blue) range of \citet{Walborn-Fitzpatrick-1990} leads to a type O5V. More precisely, the most frequently used features for determining the spectral class are HeII~$\lambda$4541 and HeI~$\lambda$4471. In our spectrum, the former is clearly stronger than the latter, pointing to a spectral type earlier than O7 (see close-up in Fig.~\ref{phd_Ostar}). This is confirmed by the comparison of HeII~$\lambda$4200 and HeII + HeI~$\lambda$4025. We also measured the equivalent width of the lines HeII~$\lambda$4541 and HeI~$\lambda$4471 to compute their ratio. Following the spectral classification relation from \citet{Mathys-1988}, we obtain $\log [EW_{4471}/EW_{4542}] = -0.48, -0.42$ depending on the continuum position, with a preference for the first value. This corresponds to spectral types O5 and O5.5 respectively. Nonetheless, given the noisy nature of the spectrum, even a spectral type O6 cannot be excluded. As for the luminosity class, the extracted absorption lines are relatively broad, with no trace of SiIV beside H$\delta$ on the higher-resolution EMMI spectra, hence broadly suggestive of a V-III luminosity class. Moreover, the line HeII~$\lambda$4686 in Fig.~\ref{phd_Ostar} is (particularly) strong in absorption, and this defines a luminosity class V.

\begin{figure}[ht]
\centering
\includegraphics[width=\linewidth]{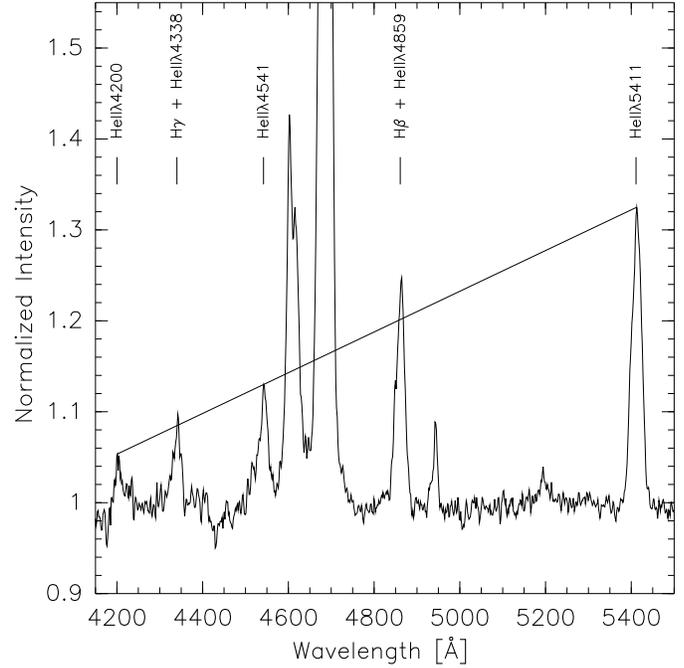}
\caption{WR spectrum obtained by a separation process on the spectra of dataset 1. The line is used to show the Pickering emission-line decrement, leading to a moderate (h) presence of hydrogen in the WN spectral type.}
\label{pickering}
\end{figure}

The WN spectrum resulting from the separation loop can be compared to the criteria of \citet*{Smith-etal-1996}. Using the peak values, we obtain a ratio for HeII~$\lambda$5411/HeI~$\lambda$5876 = 16.1. This points towards a WN3 subclass, that is confirmed by the peak ratios of NV~$\lambda$4604 and NIII~$\lambda$4640, since the latter is very weak. The practical absence of the CIV~$\lambda$5808 line makes the result definitive: WN3. Finally, we measured the Pickering emission-line decrement. Using the description of \citet{Smith-etal-1996}, the heights of H+He lines clearly exceed the line drawn between pure Helium lines. Results are shown in Fig.\,\ref{pickering}. Moreover, the two hydrogen criteria of Smith et al. are equal to 0.22 and 0.24, respectively. These are probably lower limits since the H-containing blends are more suppressed by stronger absorption components than the pure He lines. Combined with the presence of H-absorption lines, this leads to the moderate presence of hydrogen and a final type of WN3(h)a. Therefore, BAT129 appears to be an early-type hydrogen-containing WN star with a dwarf O5 companion.

\subsection{Luminosity ratio}

Using the methods described in \citet{Cherepashchuk-etal-1995} we derived the(visual) luminosity ratio $q \equiv L_{WR}/L_{O}$. To achieve a meaningful ratio, we combined EMMI spectra between phases 0.11 and 0.25, i.e. outside the eclipses phases (the spectra of dataset~1 were not of sufficient quality and resolution for this). The classical method by \citet{Beals-1944} provides an upper limit, and the new method by \citet{Cherepashchuk-etal-1995} gives the "correct" value (i.e. corrected by the fact that the absorption and emission lines are superimposed on each other). The determination of the luminosity ratio is made by comparing the equivalent widths (EW) of O-star absorption lines both in the original mean WR+O spectrum and the "pure" O-star mean spectrum. We measured the EWs on the following lines, by gaussian fitting: H$\epsilon$, HeII~$\lambda$4025, H$\delta$ + HeII~$\lambda$4100, HeII~$\lambda$4200 and H$\gamma$ + HeII~$\lambda$4338.

Two problems arise here. First, the absorption lines belonging to the WR star in the WR+O spectrum must be removed. To account for this, we corrected the EW of absorption lines in the original WR+O spectra by measuring the EW of the WR absorption lines in the extracted WR spectra:
\begin{equation}
EW_{bin} \equiv EW_{WR+O} - EW_{abs,WR}.
\end{equation}
Also:
\begin{equation}
q = \frac{EW_{single}}{EW_{bin} \cdot LCR} - 1,
\end{equation}
where LCR stands for "Line-to-continuum ratio" and is the measure of the line strength to the continuum level (i.e. peak value) in the WR+O spectrum \citep[see][ for the detailed equations]{Cherepashchuk-etal-1995}. 

The second problem is to compare our values of $EW_{single}$ to that of single O stars. We first measured the EW on the extracted O-star spectrum. Then, comparing our results with the values of \citet{Conti-Frost-1977} for the lines HeII~$\lambda$4025, HeII~$\lambda$4200 and $H\gamma$, it seems that our values for the helium lines are overestimated, while that of H$\gamma$ is underestimated (by about 0.3 and 0.6 \AA\ respectively). Taking lower values for the Helium lines give inconsistent results, since the EW on the combined binary spectrum would be larger than that of the single spectrum. Taking the upper value for H$\gamma$ would actually correct the result of this specific line to a level comparable to that of the other lines. Although not fully reliable (since the result depends heavily on the output of a complicated separation process), we consider that our measurements on the extracted spectra are nevertheless globally consistent.

Assuming of course that there is no third-light contamination, and taking these values with care, we obtain the following mean luminosity ratios: 0.39 $\pm$ 0.18 and 0.28 $\pm$ 0.20, respectively with the two methods. In the following, we assume a luminosity ratio of 0.3.

\subsection{The O-star characteristics}

Given the spectral type we assigned to the O star, the temperature-spectral~type relation published by \citet{Massey-etal-2005} in their detailed study of LMC O stars, implies an effective temperature of about 40 kK, interpolating roughly the values between the SMC and the Milky Way relations. Similarly, using the scale for a solar metallicity and for galactic dwarf stars by \citet{Martins-etal-2005} implies $T_{eff} = 41$ kK for a spectral type O5V. However, these latter authors emphasize metallicity effects which provide higher temperature at lower metallicity. If we consider a shift of about one spectral subclass, this gives (for an O4V star): $T_{eff} = 43$ kK. For such a temperature, the luminosity of the O star is $\log(L/L_{\odot}) = 5.49$. Let's assume this value of $L$ for the secondary star in BAT129.

One of the stars in the sample of \citet{Massey-etal-2005}, namely \object{LH 58-496}, has the same spectral type (O5V) and thus the assumed luminosity ($log(L/L_{\odot})=5.49$) of the O star in BAT129. The radius of the star quoted by the authors is 10.5 $R_{\odot}$. Following \citet{Eggleton-1983}, this value is slightly smaller than that of the Roche radius of the O star in BAT129 (11.8 $R_{\odot}$). Moreover, the evolutionary mass of \object{LH 58-496} is 40 $M_{\odot}$, compared to the measured Keplerian mass of BAT129: $M_O \, sin^3 i = 23 M_{\odot}$ (see Table \ref{finalorbit-tab}). To reach a value of 40 $M_{\odot}$ for BAT129, the inclination angle of BAT129 should be 57 degrees. Given the strong eclipsing nature of BAT129, this value of the inclination should be considered as a minimum estimate. Considering the large radius of the O5 star, the secondary component of BAT129 well may be an O6, with a smaller radius.

In their sample, \citet{Massey-etal-2005} have an O6.5V star, \object{AV 446} (in the SMC though). It has a temperature of $T_{eff} = 41.0$ kK, an evolutionary mass of 33 $M_{\odot}$, and a radius of 8.8 $R_{\odot}$ (i.e. 75\% of the O-star Roche radius). These values are also quite consistent with the estimates we draw on the secondary star of BAT129. 

Given the short period, it is likely that the system is synchronized. If so, then with a radius of 8.8 $R_{\odot}$, the equatorial velocity becomes 161 km~s$^{-1}$. However, the FWHM measured on the extracted O-star spectrum gives a {\it projected} rotational velocity slightly larger by about 20 to 40 km~$^{-1}$, as shown in Table \ref{fwhm_Ostar} (assuming that the dominant broadening cause is the Doppler effect). On the other hand, these values are very close to the synchronized velocity of the star if it had a radius of 10.5 $R_{\odot}$ (192 km~$^{-1}$). Oversynchronicity is already known to occur in some WR+O binaries, such as \object{V444 Cyg} \citep{Marchenko-etal-1994} and \object{CQ Cep} \citep{Marchenko-etal-1995}. 

\begin{table}[!t]
\caption{Projected rotational velocities inferred from the O-star absorption lines, assuming that the broadening of the lines is due only to Doppler broadening. Values have been corrected by the 2.6-pixel resolution.}
\label{fwhm_Ostar}
\centering
\begin{tabular}{lll}
\hline \hline
Line & FWHM (\AA) & Velocity (km~s$^{-1}$) \\ \hline
H$\epsilon$ + HeII~$\lambda$3968 		& 6.72 & 195 \\
HeII~$\lambda$4025 + HeI~$\lambda$4026 	& 5.51 & 158 \\
H$\delta$ + HeII~$\lambda$4100			& 7.43 & 209 \\
HeII~$\lambda$4200						& 6.61 & 181 \\
H$\gamma$ + HeII~$\lambda$4338			& 7.19 & 191 \\
\hline
\end{tabular}
\end{table}

\subsection{The WR-star characteristics}

The visual apparent magnitude of BAT129 quoted by \citet{bat99} is 14.92, refering to a work by \citet{Rousseau-etal-1978}. Assuming a distance modulus of 18.5 for the LMC \citep{vandenBergh-2000} and IS visual extinction of Av = 0.4 \citep[i.e. 3.1 * 0.13, see][]{Massey-2003,Conti-Morris-1990}, the absolute visual magnitude of the system is -3.98, i.e. fainter than the magnitude of the single O6.5V star mentioned above \citep[Mv = -4.7,][]{Massey-etal-2005}. The Johnson magnitude of BAT129 given by \citet{Feitzinger-Isserstedt-1983} is 14.68, i.e Mv = -4.2, which is still fainter than the O star value\footnote{We were not able to compute the correct MACHO magnitude, following equations 1 and 2 of \citet{Alcock-etal-1999}, since some coefficient values were not accessible on the Internet, at the time of writing.}. Two obvious reasons can explain the above discrepancy. First, the measured magnitude of the system is too faint because the eclipses were not accounted for. This effect is unlikely to resolve the discrepancy, since it is smaller than 0.2 magnitude (see Fig. \ref{phot_corr}). Second, the spectral type inferred from our extraction process is wrong and a cooler spectral type must actually be assigned to the O star in BAT129. This second possibility is more plausible, although we have no definite means to verify it.

We can nonetheless compare the properties of the WR star in BAT129 with LMC WN stars studied by \citet{Hamann-Koersterke-2000}. Taking our luminosity ratio of 0.3, the WR star has a luminosity of $\log(L/L_{\odot}) = 4.97$, and $M_v$=-3.34. This absolute magnitude is nonetheless close to the average value for the Galactic WN3 stars \citep{vanderHucht-2001}, and slightly fainter than that of WR3 in our Galaxy \citep[Mv = -3.7,][]{Marchenko-etal-2004}

Given the luminosity of the WR star in BAT129 and its absolute visual magnitude, the most similar star in the selected subsample of \citet{Hamann-Koersterke-2000} is \object{BAT99 63} (Brey 52): it has a spectral type WN4, a temperature of 71 kK, 50\%  hydrogen (by mass), a radius $R_{*}$ (i.e. at a Rosseland optical depth of 20) of 4.7 $R_{\odot}$ and a mass-loss rate $\log (\dot{M}/M_{\odot}) = -5.2$. This mass-loss rate value is in agreement with the recent theoretical value computed by \citet{Vink-deKoter-2005-astroph}, for the metallicity of the LMC: -5.26, -5.49 for $\beta$ (exponent of the wind velocity law) equal to 1 and 3, respectively. In the MACHO light curve, spanning more than 10 years, we have also searched for any significant period changes. We have split the lightcurve into 10 equally long bins, and performed a Fourier analysis on each of them. Note that given the number of datapoints and their spread in time, the maximum number of bins allowing a precise analysis is not much larger than 10. We found no significant period changes, with a precision of 0.003 days over the whole timespan. However, this precision does not allow us to put any meaningful constraint on the WR mass-loss rate. Comparing the Keplerian mass of the WR component in BAT129 ($M_1 sin^3 i$ = 14 $M_{\odot}$) to the {\it actual} mass of BAT63 quoted by \citet{Hamann-Koersterke-2000} -- 22 $M_{\odot}$ -- the inclination angle in BAT129 must be 59 degrees, which is equivalent to the estimate drawn from the O star.

For the moment, let us assume that the WR star in BAT129 has followed a pure single-star evolution. The WR star must have had an initial mass above that of the current secondary star, i.e., at least 35 $M_{\odot}$. On the other hand, if we assume that the WR star is, say, in the middle of its eWNL phase (i.e. its h-rich WN phase, see below), we can use the mass loss rate quoted above to estimate the amount of mass lost by the star during the WR stage. The duration of the Helium-burning phase of a 40 $M_{\odot}$ star, following the models of rotating WR stars  at Z=0.008 of \citet{Meynet-Maeder-2005}, is 0.550 million years. Assuming the WR star in BAT129 is well described by these models, this means that the star has lost slightly less than 2 $M_{\odot}$ during the WR phase. The 60 $M_{\odot}$ model would imply an even lower value, but even the 30 $M_{\odot}$ model does not give a value larger than 3 $M_{\odot}$ lost by the WR. Taking the slightly smaller mass-loss rate of WR3 in our Galaxy \citep[$\log (\dot{M}/M_{\odot}) = -5.62$,][]{Marchenko-etal-2004} of course does not resolve the situation. This means that the mass-loss rate quoted above is certainly not valid for the WR star in BAT129 {\it as a single star}, since the estimated mass lost is too small by a factor 5 (2-3 $M_{\odot}$ compared to at least 10 $M_{\odot}$ from its initial mass). 

With the assumption that BAT129 is a non-interacting system and that no RLOF occurred in its past evolution, {\it some} of the parameters of the WR star and its companion can be relatively well explained by single-star evolution. However, the combination of the possible oversynchronous rotation of the O star, the very short period of the system and the theoretical single-star mass loss rate clearly too small seem to point toward an interacting scenario for BAT129, although the presence of hydrogen in the WR star might be contradictory at first sight. This is discussed in the next section in a broader perspective of the WN evolution with hydrogen.

\section{Discussion}

It appears that weak absorption features are superposed on the emission line profiles in the extracted WR spectra. We have shown that the features are moving with the WR star. These absorption features are blueshifted relative the main emission profiles, and the blue shift seems to increase in the lines with progressively lower Pickering (or Balmer) number. Therefore, we are tempted to conclude that these absorption features are identical to the intrinsic absorption lines seen in many early-type WN Magellanic Cloud stars and \object{WR3} in our Galaxy. However, {\it for the first time we detect them in a short-period binary}. The presence of hydrogen in an early-type WN star in a binary has important consequences that are discussed below.

\subsection{Wolf-Rayet star classification}

The presence of hydrogen in WN stars is relevant for their classification and from the point of view of their evolutionary status. Most of the late-type WN stars (WNL) have hydrogen in their winds, as they are believed to be the first WR stage after the main-sequence and/or a possible transition stage. However, a significant number of exceptions exist to this rule. Similarly, most of the early-type WN stars (WNE) are hydrogen-poor, or hydrogen-free, but again, with a significant number of exceptions. For some reasons, that are likely to be due to rotation, the exceptions are more numerous in the SMC. Five out of 11 WN stars in the SMC have a very hot spectral type (WN4, WN3) and present signatures of a large amount of hydrogen (ratio H/He by number $\sim$ 1, P. Crowther, private communication). Therefore, the ionization subclass (i.e. the temperature) is not the decisive criterion to distinguish between late and early-type WN stars. This is the reason why \citet{Foellmi-etal-2003b} proposed a complementary and evolutionary classification for
WN stars (recognizable by putting an "e" in front of the class) which is based on the hydrogen content. The main eWNL/eWNE distinction is now based on the amount of hydrogen. To discriminate the h-rich from the h-free stars, the "b" (for "broad" lines) classification criterion of the classification scheme of \citet{Smith-etal-1996} can be used. As a matter of fact, \citet{Hamann-etal-1995} have shown that the width of the emission lines is directly connected to the hydrogen content \citep[see also ][]{Smith-Maeder-1998}. Moreover, since a clear anti-correlation between the "b" and "h/(h)" spectral notations of WN stars is observed in LMC stars \citep{Foellmi-etal-2003b} -- an effect already mentioned by \citet{Smith-etal-1996} -- the "b" criterion is an "antipode" of the "h" criterion.

This classification was meant for single stars, since the hydrogen content in WN stars in binaries was not known. With BAT129, we can extend the applicability of this classification. It is rather natural to think that the same causes produce the same effects: when the lines are narrow, it means that hydrogen is present in the atmosphere of a WR star, and in some cases where the amount of H is substantial, an alterning Pickering emission-line serie and/or absorption lines may be visible in the optical spectrum. Moreover, we note that this classification implies that the SMC contains virtually no eWNE stars, but only eWNL stars (and one WC/WO binary). This fact is well reproduced by the recent models of rotating WR stars at low metallicity of \citet[][ see their Fig. 9]{Meynet-Maeder-2005}.

\subsection{Evolution of WR+O binaries}

A simple estimate of the minimum period of BAT129 allows us to show that the system was certainly interacting in the past. Let us assume a ZAMS mass of the O star similar to the estimated mass above (say, 35 $M_{\odot}$, i.e. the star is assumed to have lost no mass), and the corresponding Main Sequence (MS) radius: $\sim$10$R_{\odot}$ \citep[see e.g.][ their Table 1]{Martins-etal-2005}. Assuming that the present WR star had an initial mass of, say, 45 $M_{\odot}$, and became a WR star by simple wind mass-loss, its maximum radius during the MS phase is the radius of a 45 $M_{\odot}$ MS supergiant: $\sim$20$R_{\odot}$ \citep[i.e. not considering a red supergiant nor a Luminous Blue Variable phase; see][ their Table 3 for the radius value]{Martins-etal-2005}. Assuming that the minimum distance between the two stars is at least the sum of their radii ($d_{min} > 30 R_{\odot}$), and using Kepler's third law, the minimum period of the system is 2.2 days, which is very similar to the current period. Therefore, given the extreme example considered here, the widening of the orbit from 2.2 to 2.7 days caused by a simple mass-loss by the stellar wind during the WR phase of the primary component can be ruled out. This implies that to obtain the present short period, the system must have gone through an interacting phase that shrank the orbit.

Furthermore, it is very interesting to compare the actual parameters of BAT129 with the latest models of binary star evolution of \citet*{Petrovic-etal-2005b}. The models of binary stars that lead to WR+O systems show that Roche-lobe Overflow (RLOF) and mass transfer occur for short-period systems. We can distinguish Case A mass transfer, that starts during the core hydrogen-burning phase of the primary, and Case B mass transfer starting during the shell hydrogen-burning phase. As emphasized by \citet{Petrovic-etal-2005b}, the net effect of a Case A (that also contains an intermediate Case AB mass transfer episode) or a Case B scenario is the increase of the period. If such mass transfer occurred in the past in BAT129, this means that the initial period would have been even smaller than about 2 days, which is very unlikely. Therefore, following Petrovic and collaborators, the evolution of BAT129 should be considered as the probable result of contact evolution.

\citet*{Wellstein-etal-2001} have studied the formation of massive contact binaries. Unfortunately, among the 74 models they have computed, none of their models actually include primary stars more massive than 25 $M_{\odot}$. If we nonetheless consider globally their results, their models also lead to an increase of the period after the RLOF (whatever its kind). We note here that an oversynchronous rotation velocity of the secondary star is known to occur in other WR+O systems, among which is the famous Galactic system \object{V444 Cyg} \citep{Marchenko-etal-1994}, \object{WR127} \citep{Massey-1981} and \object{CQ Cep} \citep{Marchenko-etal-1995}. Interestingly, while \object{CQ Cep} is supposed to have experienced a contact phase, this is not the case for \object{V444 Cyg}, which is relatively well explained by a stable mass transfer scenario, according to \citet{Petrovic-etal-2005b}. The uncertainty on the radius of the WR star in BAT129 is unfortunately too large to allow a conclusive discussion on this phenomenon related to the interaction of the two stars.

Altogether, it seems likely that BAT129 followed an interacting (possibly contact) evolution. A detailed analysis of the abundances of the O star could possibly help to probe the question of whether the companion star has accreted mass from the primary, and how much. We have obtained very high S/N FORS1 spectra at the VLT of BAT129 that could possibly allow such analysis. This is left for a forthcoming paper.

\subsection{Gamma-ray burst progenitors}

In this section we try to consider the binary star BAT129 or its WR component alone as "test particles" for the scenario of formation of Gamma-ray bursters (GRB).

Although WR stars are believed to be the progenitors of Type Ib/c supernovae associated with long-duration GRBs \citep{Galama-etal-1998}, quite successfully described by the "collapsar" model \citep{MacFadyen-Woosley-1999}, it is still very difficult to answer the question: what {\it kind} of WR stars are the best candidates for a GRB? \citet*{Petrovic-etal-2005a} try to address this issue from a theoretical point of view. They emphasize that three essential ingredients are needed to produce a GRB: a massive core to form a black hole (BH), sufficient angular momentum to form an accretion disc around the BH with bipolar jets, and loss of hydrogen envelope to let bipolar jets reach the star surface. They show that both single and binary non-magnetic WR stars can retain enough angular momentum for a GRB to form. Hence the question remains: are binaries more favorable GRB progenitors than single stars?

The recent models of single rotating WR stars at different metallicities of \citet{Meynet-Maeder-2005} show 3 important things. (1) Final surface velocity at low metallicity is higher than at high metallicity, a fact that is explained as a consequence of the lower mass loss rates at lower metallicity. Therefore, low metallicity galaxies such as the SMC and the LMC are better locations to find fast rotating WR stars at the end of the He burning phase, in agreement with the results of \citet{LeFloch-etal-2003} who show that GRB occur mostly in blue sub-luminous galaxies \citep[see also][]{Prochaska-etal-2004}. (2) At a given metallicity the velocities obtained at the end of the He-burning phase are larger for smaller initial mass stars. (3) The last point is that most of the models of \citet{Meynet-Maeder-2005} at low metallicity have enough specific angular momentum in their core to be good candidates for collapsar models. Again we can ask: are binary stars better GRB candidates than single stars, even if single star WR evolution provides the necessary conditions for such an event? 

How does BAT129 fit in this picture? While the metallicity is that of the LMC, and the combination of the very early spectral type and the presence of hydrogen is probably a sign of rapid rotation \citep{Foellmi-etal-2003a,Foellmi-2004}, the main problem is to know what was really the initial mass of the WR star. If it had followed a single-star evolution, it is likely that it was initially a lower-mass star (say with an initial mass between 20 and 40 $M_{\odot}$). This is consistent with its large rotational velocity \citep[see again the corresponding model of][]{Meynet-Maeder-2005}, and the clear presence of hydrogen. However, we have shown above that the characteristics of BAT129 as a binary make it a good candidate for an interacting binary, possibly with a contact phase. We can therefore ask: could the initial mass of the WR star be very large (say above 50 $M_{\odot}$), and most of it has been lost in the interaction of the binary? Could binary evolution produce such a WR+O system where the WR star looks like its SMC counterparts? If the WR star lost a large amount of mass and the secondary did not accrete all the material, it is likely that a nebula should be seen around BAT129. We find however no clear nebular lines in our spectra. 

\citet{Tutukov-Cherepashchuk-2004} also mention that the creation of a BH and the generation of a GRB associated with it is most probable in the closest massive binary systems (not only WR+O but also WR+[compact companion] binaries), since the presence of a close companion limits the rotational angular velocity of the WR core to the orbital angular velocity of the binary system. Very short period WR+O binaries formed through a common envelope phase, as BAT129 could be, is very relevant in this picture \citep[see also the models by][]{Wellstein-Langer-1999}. Although not a definitive argument, we note that the rarity of very close WR binary systems might help the frequencies of GRBs and fast rotating WR core collapses to agree \citep[see e.g.][]{Postnov-Cherepashchuk-2001}.

The point of GRB progenitors is further investigated by \citet*{Hirschi-etal-2005-astroph} for single star evolution. They note in particular that angular momentum at the end of the He-burning is a good approximation of the angular momentum contained in the core at the collapse. This is an important point for observational tests, since even the pre-core collapse evolution phases are rather short and unlikely to be extensively observed. The authors also give a rough estimate of the initial mass range to form a BH, that is between 20 and 60 $M_{\odot}$. But forming a BH is not sufficient to produce a GRB; the angular momentum must be large enough to build up an accretion disk that will power bipolar jets. Interestingly, they show that WR stars of the WO subtype are the most probable GRB progenitors, since they are formed at low metallicity, and always form a BH \citep[see also][]{Postnov-Cherepashchuk-2001}. However, the initial mass of a WO star is estimated to be $\gtrsim 50 M_{\odot}$. 

All the above considerations lead to the reasonable conclusion that GRB progenitors are certainly a special trade-off of the different parameters of WR evolution. Some things seem however to be clear. The metallicity should be low, and the initial mass of the WR star not too high to avoid any strong mass loss, and in particular any Luminous Blue Variable (LBV) event \citep[see][ for limiting masses of the 3 main evolution scenarii of WR stars]{Meynet-Maeder-2005}. It is interesting to note that in the models of \citet{Meynet-Maeder-2005} at SMC metallicity (Z = 0.004), only the 60 $M_{\odot}$ model has clearly a higher rotational velocity at the end of He burning, while the lowest initial mass models (30 and 40 $M_{\odot}$) at LMC metallicity (Z = 0.008) have the largest rotational velocity. We may wonder whether only single WO stars at the lowest metallicity produce GRBs, while lower mass WR stars do so at slightly higher metallicity. In that context, the unique WO binary star of the SMC (WR8) deserves probably much more attention \citep[see e.g.][]{Bartzakos-etal-2001}.

One may also want to discuss the growing amount of evidence that hydrogen is clearly detected in GRB afterglows. This has been first discussed by \citet{Rigon-etal-2003} for supernova SN 1999E possibly associated with the BATSE event of GRB 980910. Recently, \citet{Starling-etal-2005a} found the clear signature of hydrogen in the spectrum of the early afterglow of GRB 021004. Finally, \citet{vanMarle-etal-2005-astroph} have made evolutionary models of circumstellar material around a massive star that produces a GRB. It is therefore relevant to search for the characteristics of hydrogen in WN stars. As suggested by the analysis of WR3 in our Galaxy \citep{Marchenko-etal-2004}, the absorption lines form relatively close to the stellar core. It is likely that the absorbing material is then gradually accelerated to the terminal wind velocity. Since we have shown that it is not unusual to see a significant amount of hydrogen in single and binary WN stars at low metallicity, it might be not too surprising to observe high-velocity hydrogen absorption features in GRB afterglows, such as described by \citet{Starling-etal-2005a} on GRB 021004.

On the other hand, \citet{vanMarle-etal-2005-astroph} argue that the presence of an intermediate velocity component in the afterglow of GRB 021004 implies that the WR phase was short, i.e. that the WR shell was still intact when the star exploded. This implies that the initial mass of the progenitor must have been small (i.e. about 25 $M_{\odot}$), since the smaller the initial mass, the shorter the WR phase \citep[see Fig. 9 in][]{Meynet-Maeder-2005}. But such stars have a very short H-free WC stage. This is possibly in contradiction with the requirement of no hydrogen left in the atmosphere \citep{MacFadyen-Woosley-1999}.

It is interesting to note that theoretical models seem to produce two very different types of GRB progenitors. On one hand, \citet{Hirschi-etal-2005-astroph} show that single WO type stars are the best GRB candidates, while the presence of hydrogen in GRB afterglows and the results of \citet{vanMarle-etal-2005-astroph} point toward lower mass WR stars. Could these latter stars still produce GRB if they belong to an interacting binary system, where the companion helps to remove the hydrogen layers? The question remains open.

\section{Conclusion} 

We have presented new medium-resolution NTT-EMMI spectra of the LMC eclipsing Wolf-rayet binary BAT99-129. We have been able to separate the spectra and find a reliable orbital solution. We assign a spectral type WN3ha + O5V to the system. The O star seems to appear possibly oversynchronous with the orbital motion. We find that the WR star contains a detectable amount of hydrogen, seen in absorption in its spectrum. While its {\it actual} characteristics are shared with the eWNL stars seen in the SMC \citep{Foellmi-etal-2003a,Foellmi-2004}, the evolution that led to the present state is certainly much different. Interacting and binary models that could be applied to systems like BAT129 (hence also to V444 Cyg) would be of major interest.

BAT99-129 appears to be a very interesting piece of the puzzle of GRB progenitors. However, the most important questions remain: What are the kinds of the progenitors of the long soft GRB progenitors? Are they single or binary stars? Are they low mass or high mass WR stars? The presence of absorption lines in BAT129 proved to be of the greatest interest and further studies of GRB afterglows will certainly confirm the importance of these absorptions.

BAT129 is definitely a very interesting target for X-ray observations, considering the eclipsing nature of the system and the clear signs of wind-wind collision. Unfortunately, we found no XMM, ROSAT, nor Chandra observations of this system. Moreover, the eclipses deserve much more study. In particular, the eclipses can be used to probe the actual sizes, in physical units, of the line emitting regions. This is now possible thanks to the complete orbital solution. We obtained VLT FORS1 spectra for this purpose and their analysis will lead to a forthcoming paper.

All in all, \object{BAT99-129} appears to be very similar to the well known WR binary in our Galaxy, \object{V444 Cyg}.

\begin{acknowledgements}
CF would like to warmly thank Olivier R. Hainaut for help. CF also thanks J.-C. Bouret for useful comments, and P. Gandhi for checking X-ray data availability. AJFM is grateful to NSERC (Canada) and FQRNT (Qu\'ebec) for financial support. This paper utilizes public domain data originally obtained by the MACHO Project, whose work was performed under the joint auspices of the U.S. Department of Energy, National Nuclear Security Administration by the University of California, Lawrence Livermore National Laboratory under contract No. W-7405-Eng-48, the National Science Foundation through the Center for Particle Astrophysics of the University of California under cooperative agreement AST-8809616, and the Mount Stromlo and Siding Spring Observatory, part of the Australian National University. IRAF is distributed by the National Optical Astronomy Observatories, which are operated by the Association of Universities for Research in Astronomy, Inc., under cooperative agreement with the U.S. National Science Foundation. This research has made use of the SIMBAD database, operated at CDS, Strasbourg, France. We thank the referee, A.M. Cherepashchuk, for constructive comments, and N. Langer for suggesting to compute the minimum period.
\end{acknowledgements}

\bibliographystyle{aa}
\bibliography{/Users/cfoellmi/science/active/Litterature/mainbiblio}


\end{document}